\documentclass[twocolumn,letterpaper,aps,prl,superscriptaddress,showpacs,amsmath]{revtex4-1}

\usepackage{graphicx}

\newcommand{\vc}[1]{\boldsymbol{#1}}
\newcommand{\pd}{{\phantom{\dagger}}}

\begin{document}

\title{Doping-Induced Ferromagnetism and Possible Triplet Pairing in 
$\vc{d}^\mathbf{4}$ Mott Insulators}

\author{Ji\v{r}\'{\i} Chaloupka}
\affiliation{Central European Institute of Technology,
Masaryk University, Kotl\'a\v{r}sk\'a 2, 61137 Brno, Czech Republic}

\author{Giniyat Khaliullin}
\affiliation{Max Planck Institute for Solid State Research,
Heisenbergstrasse 1, D-70569 Stuttgart, Germany}

\begin{abstract} 
We study the effects of electron doping in Mott insulators containing $d^4$
ions such as Ru$^{4+}$, Os$^{4+}$, Rh$^{5+}$, and Ir$^{5+}$ with $J=0$ singlet
ground state. Depending on the strength of the spin-orbit coupling, the
undoped systems are either nonmagnetic or host an unusual, excitonic magnetism
arising from a condensation of the excited $J=1$ triplet states of $t_{2g}^4$.
We find that the interaction between $J$-excitons and doped carriers strongly
supports ferromagnetism, converting both the nonmagnetic and antiferromagnetic
phases of the parent insulator into a ferromagnetic metal, and further to a
nonmagnetic metal. Close to the ferromagnetic phase, the low-energy spin
response is dominated by intense paramagnon excitations that may act as
mediators of a triplet pairing.
\end{abstract}

\date{\today}




\maketitle

A distinct feature of Mott insulators is the presence of low-energy magnetic
degrees of freedom, and their coupling to doped charge carriers plays the
central role in transition metal compounds~\cite{Ima98}. In large spin systems
like manganites, this coupling converts parent antiferromagnet (AF) into a
ferromagnetic (FM) metal and gives rise to large magnetoresistivity effects.
The doping of spin one-half compounds like cuprates and titanites, on the other
hand, suppresses magnetic order and a paramagnetic (PM) metal emerges. In
general, the fate of magnetism upon charge doping is dictated by spin-orbital
structure of parent insulators.  
 
In compounds with an even number of electrons on the $d$ shell, one may
encounter a curious situation when the ionic ground state has no magnetic
moment at all, yet they may order magnetically by virtue of low-lying magnetic
levels with finite spin, if the exchange interactions are strong enough to
overcome single-ion magnetic gap. The $d^4$ ions such as Ru$^{4+}$, Os$^{4+}$, 
Rh$^{5+}$, Ir$^{5+}$ possess exactly this type level structure~\cite{Abr70} 
due to spin-orbit coupling $\lambda(\vc{S}\!\cdot\!\vc{L})$: 
the spin $S=1$ and orbital $L=1$ moments form a
nonmagnetic ground state with total $J=0$ moment, separated from the excited
level $J=1$ by $\lambda$. A competition of the exchange and spin-orbit couplings
results then in a quantum critical point (QCP) between nonmagnetic Mott
insulator and magnetic order~\cite{Kha13,Mee15}. Since magnetic order is due
to condensation of the virtual $J=1$ levels and hence ``soft'', the amplitude
(Higgs) mode is expected. The corollary of the ``$d^4$ excitonic
magnetism''~\cite{Kha13} is the presence of magnetic QCP that does not require
any special lattice geometry, and the energy scales involved are large. The
recent neutron scattering data~\cite{Jai15} in $d^4$ Ca$_2$RuO$_4$ seem to
support the theoretical expectations. 

As we show in this Letter, unusual magnetism of $d^4$ insulators, where 
the ``soft'' $J$-spins fluctuate between 0 and 1, results also in anomalous
doping effects that differ drastically from conventional cases as manganites 
and cuprates. Indeed, while common wisdom suggests that the PM phase with
yet uncondensed $J$-moments near QCP would get even ``more PM'' upon doping, 
we find that mobile carriers induce long-range order instead. The order is of
FM type and is promoted by carrier-driven condensation of $J$-moments. 
By the same mechanism, the exchange dominated AF phase also readily switches 
to FM metal, as observed in La-doped Ca$_2$RuO$_4$~\cite{Cao00,Cao01}.
The theory might be relevant also to electric-field-induced FM of
Ca$_2$RuO$_4$~\cite{Nak13} and FM state of the RuO$_2$ planes in oxide
superlattices~\cite{Hug15}. Further doping suppresses any magnetic order, 
and we suggest that residual FM correlations may lead to a triplet 
superconductivity (SC). 

\begin{figure}[htb] 
\includegraphics[width=7.7cm]{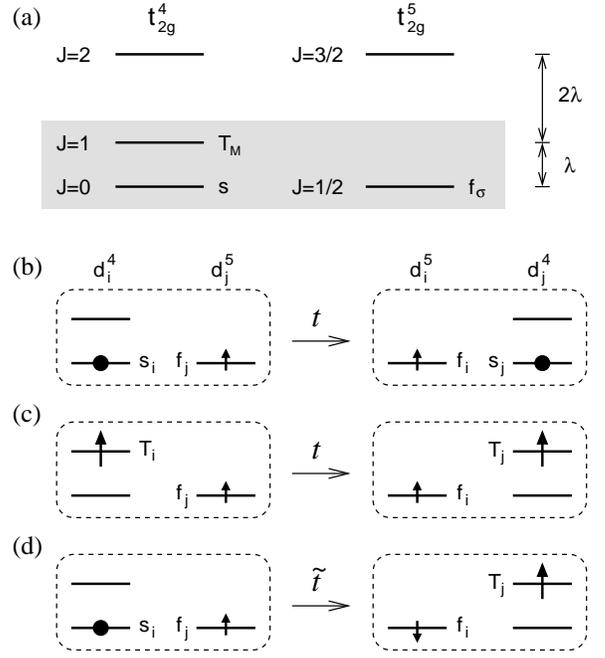} 
\caption{
(a)~Spin-orbital level structure of $t_{2g}^4$ and $t_{2g}^5$ configurations. 
Lowest states including singlet $s$ and triplet $T_M$ states of $d^4$, and 
pseudospin $1/2$ $f_\sigma$ states of $d^5$ configurations form a basis for 
effective low-energy Hamiltonian.
(b)-(d)~Schematics of electron hoppings that lead to Eqs.~\eqref{eq:h1} 
and \eqref{eq:h2}: (b)~Free motion of a doped fermion $f_\sigma$ in a singlet
background.
(c)~Fermion hopping is accompanied by a triplon backflow supporting
double-exchange type ferromagnetism.
(d)~Fermionic hopping generates a singlet-triplet excitation. This process 
leads to a coupling between Stoner continuum and $T$-moments promoting
magnetic condensation.
}
\label{fig:schem}
\end{figure}

{\it Model}.-- There are a number of $d^4$ compounds, magnetic as well
nonmagnetic, with various lattice
structures~\cite{Nak97,Miu07,Bre11,Cao14,Shi13,Kha02,Lee06,Hua06}. To be
specific, we consider a square lattice $d^4$ insulator lightly doped by
electrons. Assuming relatively large spin-orbit coupling (SOC), the relevant 
states are pseudospin $J=0,1$ states of $t_{2g}^4$ and
$J=1/2$ states of $t_{2g}^5$ [see Fig.~\ref{fig:schem}(a)]. The $d^4$ singlet
$s$ ($J=0$) and triplon $T_{0,\pm 1}$ ($J=1$) states obey the Hamiltonian
derived in Ref.~\onlinecite{Kha13}. Adopting the Cartesian basis 
$T_x\!=\!(T_1\!-\!T_{-1})/\sqrt{2}i$, $T_y\!=\!(T_1\!+\!T_{-1})/\sqrt{2}$, and 
$T_z\!=\!i T_0$, it can be written as 
\begin{multline}
\mathcal{H}_{d^4} = 
\lambda \sum_i \vc T_{\!i}^\dagger \cdot \vc T_{\!i}^\pd
+\tfrac14 K \sum_{\langle ij\rangle} \left[
s^\pd_i s_j^\dagger (\vc T_{\!i}^\dagger \cdot \vc T_{\!j}^\pd 
\!-\!\tfrac13 T_{i\gamma}^\dagger T_{j\gamma}^\pd) \right. \\ \left.
-s_i^\dagger s_j^\dagger(\tfrac56 \vc T_{\!i}\cdot \vc T_{\!j} 
\!-\!\tfrac16 T_{i\gamma} T_{j\gamma}) +\mathrm{H.c.}\right] ,
\end{multline}
where $\gamma$ is determined by the bond direction. The model shows 
AF transition due to a condensation of $\vc T$ at a critical value
$K_c=\frac{6}{11}\lambda$ of the interaction parameter $K=4t_0^2/U$. 
The degenerate $T_{x,y,z}$ levels split upon material-dependent lattice
distortion, affecting the details of the model behavior~\cite{Akb14}. We will
consider the cubic symmetry case and make a few comments on the possible effects
of the tetragonal splitting.   

The $d^4$ system is doped by introducing a small amount of $d^5$ objects --
fermions $f_{\sigma}$ carrying the pseudospin $J=1/2$ of $t_{2g}^5$. The
on-site constraint $n_s+n_T+n_f=1$ is implied. The Hamiltonian describing the
correlated motion of $f$ is derived by calculating matrix elements of the
nearest-neighbor hopping $\hat T_{ij}=-t_0(a^\dagger_{i\sigma}a^\pd_{j\sigma}
+b^\dagger_{i\sigma}b^\pd_{j\sigma})$ between multielectron configurations 
$\langle d_i^5d_j^4|\hat T_{ij}|d_i^4d_j^5\rangle$.  
Here $a$ and $b$ are the $t_{2g}$ orbitals active on a given bond,
\textit{e.g.} $xy$ and $zx$ for $x$-bonds. The resulting hopping Hamiltonian
comprises three contributions,
$\mathcal{H}_{d^4\textrm{-}d^5}=\sum_{ij}(h_1 + h_2 + h_3)_{ij}^{(\gamma)}$.
The first one, depicted schematically in Fig.~\ref{fig:schem}(b,c), 
is a spin-independent motion of $f$, accompanied by a backflow of 
$s$ and $\vc T$:
\begin{equation}\label{eq:h1}
h_1^{(\gamma)}\!=\!-t f_{i\sigma}^\dagger f_{j\sigma}^\pd
\left[s_j^\dagger s_i^\pd+\tfrac{15}{16}(\vc T_{\!j}^\dagger\cdot\vc T_{\!i}^\pd
\!-\!\tfrac35 T_{j\gamma}^\dagger T_{i\gamma}^\pd)\right].
\end{equation}
The second contribution is a spin-dependent motion of $f$ generating 
$J\!=\!0$ $\leftrightarrow$ $J\!=\!1$ magnetic excitation in the 
$d^4$ background [see Fig.~\ref{fig:schem}(d)]: 
\begin{equation}\label{eq:h2}
h_2^{(\gamma)}\!=\! i\tilde{t} \left[
\sigma_{ij}^\gamma (s_j^\dagger T_{i\gamma}^\pd
                    \!-\!T_{j\gamma}^\dagger s_i^\pd)
\!-\!\tfrac13
\vc{\sigma}_{ij}\!\cdot\! (s_j^\dagger \vc T_{\!i}^\pd
                    \!-\!\vc T_{\!j}^\dagger s_i^\pd)
\right].
\end{equation}
Here, $\vc \sigma_{ij}=f_{i\alpha}^\dagger \vc{\tau}_{\alpha\beta}^\pd 
f_{j\beta}^\pd$ with Pauli matrices $\tau$ denotes the bond-spin operator.
The derivation for the cubic symmetry gives $t=\tfrac49 t_0$ and
$\tilde{t}=\frac1{\sqrt{6}}\,t_0$ with the ratio $\tilde{t}/t\approx 1$.
However, these values are affected by the lattice distortions (via the
pseudospin wave functions) and $f$-band renormalization reducing the effective
$t$. We thus consider $\tilde{t}/t$ as a free parameter and set
$\tilde{t}=1.5t$ below. The last contribution to
$\mathcal{H}_{d^4\textrm{-}d^5}$ reads as coupling between the bond-spins
residing in $f$ and $T$ sectors: $h_3^{(\gamma)}= 
\frac{9}{16}t\left(\sigma_{ij}^\gamma J^\gamma_{ji} 
+\tfrac13 \vc{\sigma}_{ij}\cdot \vc J_{ji} \right)$, 
where $\vc J_{ji}=-i(\vc T_{\!j}^\dagger \times \vc T_{\!i}^\pd)$.  
At small doping and near QCP where the density of $\vc T$ excitons is small,
the scattering term $h_3$ can be neglected.

{\it Phase diagram}.-- We first inspect the phase behavior
of the model as a function of doping $x$ and interaction parameters $K$ and
$\tilde{t}$. The magnetic order is linked to the condensation of triplons
induced by their mutual interactions and the interaction with the doped
fermions $f$. In contrast to the cubic lattice where all the $\vc T$
flavors are equivalent, on the two-dimensional square lattice the $T_z$
flavor experiences the strongest interactions and is selected to condense,
provided that it is not suppressed by a large tetragonal distortion. We thus
focus on $T_z$ and omit the index~$z$.

Following the standard notation for spin-1 condensates, we express complex
$\vc T=\vc u+i\vc v$ using two real fields $\vc u$, $\vc v$. The ordered
dipolar moment residing on Van Vleck transition $s\leftrightarrow \vc T$ is
then $\vc m=2\sqrt{6}\,\vc v$~\cite{Kha13}. Assuming either FM order
(condensation prescribed by $T \rightarrow iv$) or AF order ($T \rightarrow
\pm iv$ in a N\'eel pattern), we evaluate the classical energy of the
$T$-condensate and add the energy of the $f$-bands polarized due to the
condensed $T$. Doing so, we replace $s_i$ by $\sqrt{1-x-v^2}$ to incorporate
the constraint on average. The resulting total energy $E(v)=E_T +
E_\mathrm{band}$ is minimized with respect to the condensate strength $v$ and
compared for the individual phases: FM, AF, and PM ($v=0$). The condensate
energy amounts to $E_T=[\lambda\pm\frac{11}6K(1-x-v^2)]v^2$, with the $+/-$
sign for FM/AF phase, respectively. The band energy
$E_\mathrm{band} = \sum_{\vc k\sigma} \varepsilon_{\vc k\sigma} n_{\vc k\sigma}$
is calculated for a particular doping level 
$x=\sum_{\vc k\sigma} n_{\vc k\sigma}$ using the band dispersion 
$\varepsilon_{\vc k\sigma}=-4(t_1-\sigma t_2)\gamma_{\vc k}$
where $\gamma_{\vc k}=\frac12(\cos k_x+\cos k_y)$. The hopping parameter 
$t_1$ stemming from $h_1$ reads as $t_1\simeq t(1-x)$ and 
$t_1\simeq t(1-x-2v^2)$ for FM and AF, respectively. This captures the 
double-exchange nature of $h_1$ -- only FM-aligned $T$ allow for a free 
motion of $f$, while AF order of $T$ blocks it. The parameter $t_2$ 
quantifies the polarization of the bands by virtue of $h_2$ and is nonzero 
in FM case only: $t_2=\frac23 \tilde{t} v \sqrt{1-x-v^2}$.

\begin{figure}[tb] 
\includegraphics[width=8.3cm]{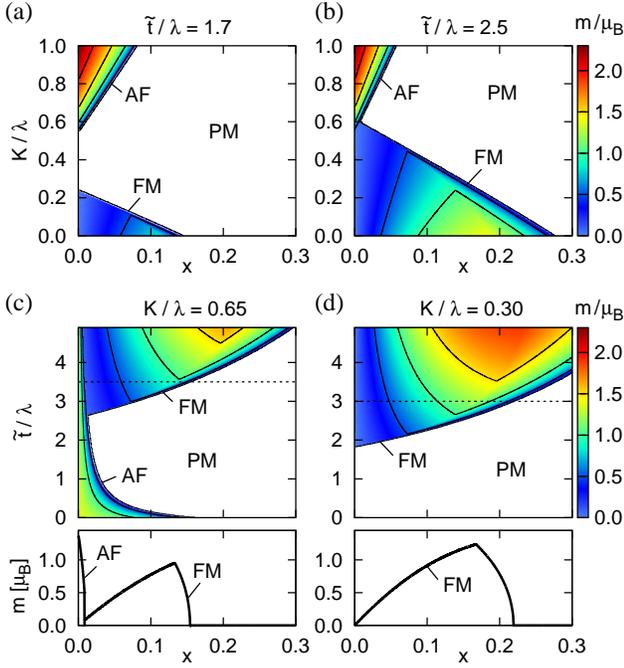} 
\caption{(Color online)
(a,b)~Phase diagrams and the ordered magnetic moment value for varying 
doping $x$ and $K/\lambda$ keeping fixed $\tilde{t}/\lambda$ of $1.7$ 
and $2.5$. 
(c)~Phase diagram for varying doping and $\tilde{t}/\lambda$ and fixed
$K=0.65\lambda$ above the critical $K_c=\frac6{11}\lambda$ of the $d^4$ 
system. Bottom panel shows $m(x)$ along the cut at $\tilde{t}/\lambda=3.5$.
(d)~The same for $K=0.3\lambda$ and the cut at $\tilde{t}/\lambda=3$.
}
\label{fig:PD}
\end{figure}

Shown in Fig.~\ref{fig:PD} are the resulting phase diagrams along with the
total ordered moment $m[\mu_B]=2\sqrt{6}\,v + n_\uparrow-n_\downarrow$. In
both phase diagrams for constant~$\tilde{t}/\lambda$ [Fig.\ref{fig:PD}(a,b)]
at $x=0$ we recover the QCP of the $d^4$ model. Nonzero doping causes a
suppression of the AF phase via the double-exchange mechanism in $h_1$, and an
appearance of FM phase strongly supported by $h_2$ that directly couples the
moment $\vc m\sim\vc v$ of $\vc T$ exciton to the fermionic spin 
$\vc \sigma_{ij}$, promoting magnetic condensation.
With increasing $\tilde{t}$ the FM phase quickly extends as seen also in
Fig.~\ref{fig:PD}(c,d) containing the phase diagrams for constant
$K/\lambda=0.65$ (selected to roughly reproduce experimental value
$1.3\:\mu_B$ for Ca$_2$RuO$_4$ \cite{Bra98}) and $K/\lambda=0.30$. The constant
$\tilde{t}/\lambda$ cut in Fig.~\ref{fig:PD}(c) is strongly reminiscent of the
phase diagram of La-doped Ca$_2$RuO$_4$ \cite{Cao00,Cao01,notepolaron}, where
the AF phase is almost immediately replaced by the FM phase present up to a
certain doping level. 
To estimate realistic values of $\tilde{t}/\lambda$, we assume
$t_0\sim 300\:\mathrm{meV}$. Large SOC in $d^4$ Ir$^{5+}$ with 
$\lambda\sim 200\:\mathrm{meV}$~\cite{Fig00,Kim12,notelambda} leads to 
$\tilde{t}/\lambda \sim 1$ and places it strictly to the AF/PM (c) or PM/PM 
(d) regime. In contrast to this, moderate $\lambda\sim 70-80\:\mathrm{meV}$ 
in Ru$^{4+}$~\cite{Abr70,Miz01} makes the FM phase easily accessible.

{\it Spin susceptibility, emergence of paramagnons}.-- The tendency toward FM
ordering naturally manifests itself in the dynamic spin response of the
coupled \mbox{$\vc T$-exciton} and \mbox{$f$-band} system. Here we study it in
detail for the PM phase, focusing again on $T_z$ being the closest to
condense. The magnetic moment $\vc m$ is carried mainly by the dipolar
component $\vc v=(\vc T\!-\!\vc T^\dagger)/2i$ of triplons so that the 
dominant contribution to the spin susceptibility is given by the 
\mbox{$\vc v$-susceptibility} $\chi(\vc q,\omega)$. To evaluate it, we 
replace $s^\pd_i \rightarrow \sqrt{1-x-n_{Ti}}$, and decouple $h_1$ 
\eqref{eq:h1} into $f$ and $T$ parts on a mean-field level. This yields 
a fermionic Hamiltonian 
$\mathcal{H}_f=\sum_{\vc k\sigma} \varepsilon_{\vc k} 
f^\dagger_{\vc k\sigma} f^\pd_{\vc k\sigma}$ with 
$\varepsilon_{\vc k}=-4 t (1-x) \gamma_{\vc k}$, and 
a quadratic form for $T_z$ boson:
$\mathcal{H}_T = \sum_{\vc q} 
[ A_{\vc q} T^\dagger_{\vc q} T^\pd_{\vc q}
-\frac12 B_{\vc q} (T^\pd_{\vc q}T^\pd_{-\vc q}
+T^\dagger_{\vc q}T^\dagger_{-\vc q}) ]$. 
Here, $A_{\vc q}=\lambda+4 t\langle n_{ij}\rangle 
(1-\gamma_{\vc q})+K(1-x)\gamma_{\vc q}$,  
$B_{\vc q}=\frac56 K(1-x)\gamma_{\vc q}$, and  
$\langle n_{ij}\rangle = \sum_{\vc k \sigma}\gamma_{\vc k} n_{\vc k\sigma}$.
Bogoliubov diagonalization provides the bare triplon dispersion 
$\omega_{\vc q}=(A_{\vc q}^2-B_{\vc q}^2)^{1/2}$ and the bare $v$-susceptibility 
$\chi_0(\vc q,\omega) = \frac12 
({A_{\vc q}-B_{\vc q}})/[{\omega_{\vc q}^2-(\omega+i\delta)^2}]$.
The susceptibility is further renormalized by the coupling $h_2$
\eqref{eq:h2}, which can be viewed as an interaction between a dipolar component
$v$ of the triplons and the Stoner continuum of $f$-fermions: 
\begin{equation}\label{eq:Hint}
\mathcal{H}_\mathrm{int} = 
g \!\sum_{\vc q} v_{\vc q} \tilde{\sigma}_{-\vc q} \;, \quad
\tilde{\sigma}_{-\vc q}\!=\!\sum_{\vc k} 
\Gamma_{\vc k\vc q}^\pd
f_{\vc k+\vc q,\alpha}^\dagger \tau^z_{\alpha\beta}f_{\vc k,\beta}^\pd \;.
\end{equation}
The coupling constant $g=\frac83\tilde{t}\sqrt{1-x}$, and the vertex 
$\Gamma_{\vc k\vc q}=\frac12(\gamma_{\vc k}+\gamma_{\vc k+\vc q})$
is close to $1$ in the limit of small $\vc k$, $\vc q$. 
By treating this coupling on a RPA level, we arrive at the full 
\mbox{$v$-susceptibility} $\chi=\chi_0/(1-\chi_0 \Pi)$ with the $v$-selfenergy
\begin{equation}
\Pi(\vc q,\omega) = g^2 \sum_{\vc k\sigma} 
\Gamma_{\vc k\vc q}^2\,
\frac{n_{\vc k\sigma}-n_{\vc k+\vc q\sigma}}
{\varepsilon_{\vc k+\vc q}\!-\!\varepsilon_{\vc k}\!-\!\omega\!-\!i\delta} \;.
\end{equation}

\begin{figure}[b] 
\includegraphics[width=8.5cm]{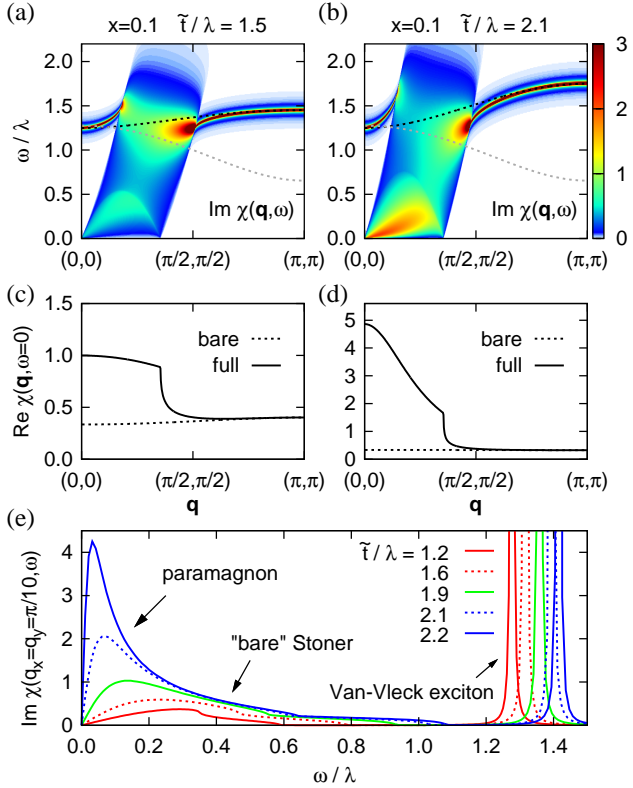} 
\caption{(Color online)
(a) Imaginary part of the \mbox{$v_z$-susceptibility} $\chi(\vc q,\omega)$ in
$(\pi,\pi)$-direction calculated for $x=0.1$, $\tilde{t}/\lambda=1.5$,
$K/\lambda=0.3$. $\chi(\vc q,\omega)$ is shown in units of $\lambda^{-1}$. 
Black (gray) dashed line shows the bare triplon dispersion for $x=0.1$ ($x=0$).
(b) The same for $\tilde{t}/\lambda=2.1$ closer to the FM transition point 
$\tilde{t}/\lambda\approx 2.25$.
(c,d) The static susceptibility corresponding to panels (a) and (b).
(e) Imaginary part of $\chi(\vc q,\omega)$ at $\vc q=(\pi/10,\pi/10)$ for 
several values of $\tilde{t}/\lambda$ gradually approaching 
the FM transition point.
}
\label{fig:susc}
\end{figure}

The interplay of the coupled excitonic and band spin responses is demonstrated
in Fig.~\ref{fig:susc}. The high-energy component of $\chi$ linked to $\chi_0$
follows the bare triplon dispersion $\omega_{\vc q}$. In an undoped system,
due to the AF \mbox{$K$-interaction}, $\omega_{\vc q}$ has a minimum at $\vc
q=(\pi,\pi)$ and $\chi_0$ would be most intense there. By doping, the double
exchange mechanism in $h_1$ disfavoring AF correlations pushes $\omega_{\vc
q}$ up near $(\pi,\pi)$.
Further, due to a dynamical mixing (\ref{eq:h2},\ref{eq:Hint}) of triplons with
the fermionic continuum, the low-energy component of $\chi$ gains spectral weight
as $\tilde{t}/\lambda$ approaches the critical value, and a gradually
softening FM-paramagnon is formed [see Fig.~\ref{fig:susc}(b)]. The emergence
of the paramagnon and the increase of its spectral weight is shown in detail
in Fig.~\ref{fig:susc}(e). Finally, once the critical $\tilde{t}/\lambda$ is
reached, triplons, whose spectral weight was pulled down by the coupling to
the Stoner continuum, condense and the FM order sets in, signaled by the
divergence of $\chi(\vc q=0,\omega=0)$ [cf.~\ref{fig:susc}(c,d)].

{\it Triplet pairing}.-- Intense paramagnons emerging in the proximity to the
FM phase may serve as mediators of a triplet pairing interaction
\cite{notepseudo}. In the following, we perform semiquantitative estimates
for this triplet SC. 

\begin{figure}[tb] 
\includegraphics[width=8.5cm]{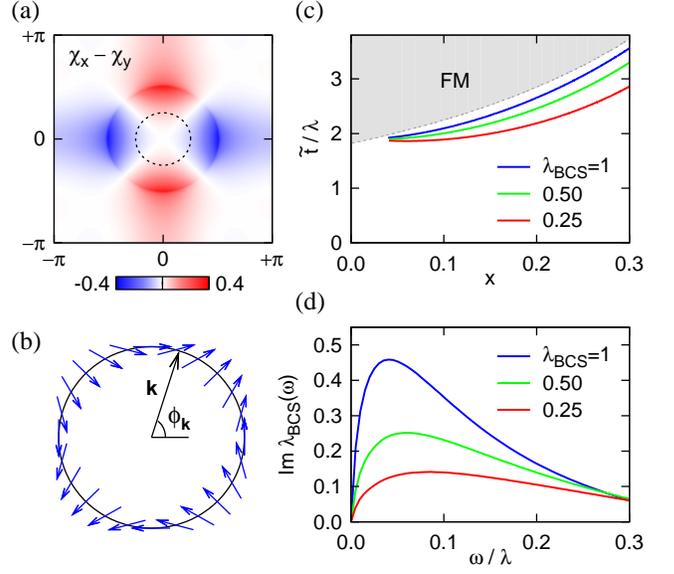} 
\caption{(Color online)
(a) Combination $(\chi_x-\chi_y)_{\omega=0}$ that determines the symmetry 
of the pairing potential $V_x-V_y$. The parameters are the same as in 
Fig.~\ref{fig:susc}(b,d). Dashed circle indicates the Fermi surface.
(b)~Representation of 
$\Delta_{\pm 1\vc k} = \Delta \mathrm{e}^{\pm i\phi_{\vc k}}$
using the \mbox{$\vc d$-vector} along the Fermi surface.
(c)~Contours of $\lambda_\mathrm{BCS}=V_0 N$ in the phase diagram of 
Fig.~\ref{fig:PD}(d).
(d)~Imaginary part of \mbox{$\omega$-dependent} $\lambda_\mathrm{BCS}(\omega)$
for $x=0.1$ and the values of $\tilde{t}/\lambda$ corresponding to 
$\lambda_\mathrm{BCS}=1$, $0.5$, and $0.25$.
}
\label{fig:SC}
\end{figure}

While the dominant contribution to the pairing strength is due to the
$v_z$-fluctuations, in order to assess the structure of the triplet order
parameter, the full coupling
$\mathcal{H}_\mathrm{int}=g\sum_{\vc q}\vc v_{\vc q}\cdot\tilde{\vc \sigma}_{-\vc q}$
leading to the effective interaction
$-\frac12g^2\sum_{\vc q\alpha}\chi_\alpha(\vc q,\omega=0) 
\,\tilde\sigma_{\vc q}^\alpha\, \tilde\sigma_{-\vc q}^\alpha$
has to be considered. The $v_\alpha$-susceptibility $\chi_\alpha$ 
for $\alpha=x,y$ may be calculated the same way as $\chi_z$ above, using now
$A_{\vc q}^\alpha=A^z_{\vc q}+
[\frac65 t\langle n_{ij}\rangle-\frac16K(1-x)]\cos q_\alpha$ and 
$B_{\vc q}^\alpha = B^z_{\vc q}-\frac1{12} K(1-x)\cos q_\alpha$.
The coupling vertex for $v_x$ and $v_y$ obtains an additional contribution, 
$\Gamma_{\vc k\vc q}^\alpha=\Gamma_{\vc k\vc q}^z
-\frac34[\cos k_\alpha +\cos (k_\alpha\!+\!q_\alpha)]$.
The resulting BCS interaction in terms of 
$t_{+1\vc k}=f_{\vc k\uparrow}f_{-\vc k\uparrow}$,
$t_{0\vc k}=\frac1{\sqrt2}(f_{\vc k\downarrow}f_{-\vc k\uparrow}
+f_{\vc k\uparrow}f_{-\vc k\downarrow})$,
and
$t_{-1\vc k}=f_{\vc k\downarrow}f_{-\vc k\downarrow}$
takes the form
\begin{multline}\label{eq:HBCS}
\mathcal{H}_\mathrm{BCS} = -\frac12 \sum_{\vc k\vc k'} \Bigl[
V_z\,(t^\dagger_{1} t^\pd_{1} \!+ t^\dagger_{-1} t^\pd_{-1})_{\vc k\vc k'}+ \\
+\!(V_x\!-\!V_y)(t^\dagger_{1} t^\pd_{-\!1}\! + t^\dagger_{-\!1}
t^\pd_{1})_{\vc k\vc k'}
+\!(V_x\!+\!V_y\!-\!V_z) t^\dagger_{0\vc k} t^\pd_{0\vc k'} \Bigr],
\end{multline}
where $V_\alpha$ denotes the properly symmetrized
$V_{\alpha\vc k\vc k'} = g^2 (\Gamma^\alpha_{\vc k,\vc k'-\vc k})^2
\frac12 [\chi_\alpha(\vc k-\vc k')-\chi_\alpha(\vc k+\vc k')]$.
Decomposed into the Fermi surface harmonics, the BCS interaction 
is well approximated by
$V_{z\vc k\vc k'}\approx 2V_0\cos(\phi_{\vc k}-\phi_{\vc k'})$ and
$(V_x-V_y)_{\vc k\vc k'}\approx 2V_1\cos(\phi_{\vc k}+\phi_{\vc k'})$
with $V_{0,1}>0$ [see Fig.~\ref{fig:susc}(d) and Fig.~\ref{fig:SC}(a)]. The
relatively small $V_1\ll V_0$ fixes the relative phase of the $t_{+1}$ and
$t_{-1}$ pairs so that the SC order parameter becomes $\Delta_{\pm 1\vc k} =
\Delta \mathrm{e}^{\pm i\phi_{\vc k}}$. This ordering type is captured by the
$\vc d$-vector $\vc d = -i\Delta(\sin\phi_{\vc k},\cos\phi_{\vc k},0) \sim
\hat{x}k_y+\hat{y}k_x$ shown in Fig.~\ref{fig:SC}(b). In the classification of
Ref.~\onlinecite{Sig91}, it forms the $\Gamma_4^-$ irreducible representation
of tetragonal group $D_{4h}$. However, this result applies to cubic symmetry 
case. Lattice distortions that cause splitting among $T_{x,y,z}$ and modify 
the pseudospin wave functions may in fact offer a possibility to ``tune'' the 
symmetry of the order parameter. If distortions favor $T_{x,y}$, the potentials
$V_{x,y}$ are expected to dominate in Eq.~\ref{eq:HBCS}, supporting the 
chiral $t_0$-pairing represented by the last term in~\eqref{eq:HBCS}.

Data in Fig.~\ref{fig:SC}(c,d) serve as a basis for a rough $T_c$ estimate.
Fig.~\ref{fig:SC}(c) shows the BCS parameter $\lambda_\mathrm{BCS}\approx V_0 N$
($N$ is DOS per spin component of the $f$-band) which attains sizable values
near the FM phase boundary, where the paramagnons are intense. To avoid
complex physics near the very vicinity of the FM QCP~\cite{Chu03,Chu04,Chu09}, 
we take a conservative upper limit $\lambda_\mathrm{BCS}\approx 0.5$.
Extending $V_0$ by the $\omega$-dependence of the underlying 
$\chi_z(\vc q,\omega)$, we define $\lambda_\mathrm{BCS}(\omega)$.
Its imaginary part to be understood as the conventional $\alpha^2 F$ 
is plotted in Fig.~\ref{fig:SC}(d) yielding an estimate of the BCS cutoff
$\Omega\lesssim 0.1\lambda$. With $\lambda\sim 100\:\mathrm{meV}$, 
this gives $T_c\approx \Omega\,\mathrm{e}^{-1/\lambda_\mathrm{BCS}}$
of about $10\:\mathrm{K}$. 

In conclusion, we have explored the doping effects in spin-orbit $d^4$ Mott
insulators. The results show that the doped electrons moving in the
$d^4$ background firmly favor ferromagnetism, explaining \textit{e.g.} the
observed behavior of La-doped Ca$_2$RuO$_4$. In the paramagnetic phase near
the FM QCP, the incipient FM correlations are manifested by intense
paramagnons that may provide a triplet pairing. 

We thank G.~Jackeli for useful comments.
J.C. acknowledges support by the Czech Science Foundation (GA\v{C}R) under 
Project No. 15-14523Y and ERDF under Project CEITEC (CZ.1.05/1.1.00/02.0068).


\begin{thebibliography}{99}
\bibitem{notepolaron} For a detailed comparison, one has to keep in mind
polaronic and phase coexistence effects \cite{Brz15} generic to weakly doped
Mott insulators, and deviations from two dimensionality.
\bibitem{notepseudo}
We recall that the pseudospins are spin-orbit entangled objects \cite{Kha05},
and hence a pseudospin triplet is in fact a mixture of real-spin singlets and
triplets.
\end{thebibliography}
\end{document}